\documentclass[pra,twocolumn]{revtex4-1}

\usepackage{amsmath}
\usepackage{amssymb}
\usepackage{graphicx}
\usepackage{mathbbol}
\usepackage{url}

\newcommand{\tr}[0]{\textrm{tr}}
\newcommand{\bra}[1]{\langle #1 |}
\newcommand{\ket}[1]{| #1 \rangle}
\newcommand{\braket}[2]{\langle #1 | #2 \rangle}

\begin{document}

\title{Optimal continuous-variable teleportation under energy constraint}

\author{Jaehak Lee}
\author{Jiyong Park}
\author{Hyunchul Nha}
\affiliation{Department of Physics, Texas A \& M University at Qatar, P.O. Box 23874, Doha, Qatar}

\begin{abstract}
Quantum teleportation is one of the crucial protocols in quantum information processing. It is important to accomplish an efficient teleportation under practical conditions, aiming at a higher fidelity desirably using fewer resources. The continuous-variable (CV) version of quantum teleportation was first proposed using a Gaussian state as a quantum resource, while other attempts were also made to improve performance by applying non-Gaussian operations. We investigate the CV teleportation to find its ultimate fidelity under energy constraint identifying an optimal quantum state. For this purpose, we present a formalism to evaluate teleportation fidelity as an expectation value of an operator. Using this formalism, we prove that the optimal state must be a form of photon-number entangled states. We further show that Gaussian states are near-optimal while non-Gaussian states make a slight improvement and therefore are rigorously optimal, particularly in the low-energy regime.
\end{abstract}

\maketitle

\section{\label{sec:introduction}Introduction}

% arising importance in CV information task
% Gaussian vs non-Gaussian
Continuous-variable (CV) systems have drawn much attention with their practical applications to quantum information processing \cite{bib:Rev.Mod.Phys.77.513,bib:Rev.Mod.Phys.84.621}, which includes quantum computation \cite{bib:Phys.Rev.Lett.97.110501, bib:Phys.Rev.A.73.032318}, quantum communication \cite{bib:Phys.Rev.A.59.1820, bib:Phys.Rev.A.63.032312, bib:Phys.Rev.Lett.92.027902, bib:Nat.Photonics.8.796, bib:Phys.Rev.A.91.042336, bib:Phys.Rev.A.93.050302}, quantum cryptography \cite{bib:Phys.Rev.Lett.88.057902, bib:Phys.Rev.Lett.101.200504}, and so on. To accomplish a given quantum task more efficiently, numerous studies aim to identify an optimal quantum state for best performance. An important line of study for CV quantum informatics is to compare performance between Gaussian and non-Gaussian states (operations). Many studies have shown that applying non-Gaussian operations such as photon subtraction, photon addition, or their coherent superposition, on Gaussian states can enhance the properties of quantum states in terms of, e.g., entanglement, teleportation fidelity, and nonlocality \cite{bib:Phys.Rev.A.61.032302, bib:Phys.Rev.A.65.062306, bib:Phys.Rev.A.67.032314, bib:Phys.Rev.A.73.042310, bib:Phys.Rev.A.80.022315, bib:Phys.Rev.A.84.012302, NhaCV,NhaCV',Park2012}.
% energy constraint to be fair comparison
However, from a resource-theoretic perspective, it is a question of interest whether such enhancement can be attributed to a more desirable property of the output quantum state or only to the consumption of more resources. For instance, the energy of a state after applying a non-Gaussian operation on it can increase while giving an enhanced fidelity of quantum teleportation \cite{bib:Phys.Rev.A.61.032302, bib:Phys.Rev.A.65.062306, bib:Phys.Rev.A.67.032314, bib:Phys.Rev.A.73.042310}. Then, it would not be fair to compare the original Gaussian state and the resulting non-Gaussian state as they do not work at the same energy level. 

In fact, one can readily anticipate that all quantum tasks in infinite dimensions can be perfectly achieved in the limit of infinite energy. In this sense, energy is a natural measure of resource to consider for a fair comparison, as one always has a finite amount of energy available in practical situations. 
Along this line, the communication capacity of bosonic channels was evaluated under the constraint of fixed energy per channel use \cite{bib:Phys.Rev.A.59.1820, bib:Phys.Rev.A.63.032312, bib:Phys.Rev.Lett.92.027902, bib:Nat.Photonics.8.796, bib:Phys.Rev.A.91.042336, bib:Phys.Rev.A.93.050302} and it was particularly shown that Gaussian states are the optimal resources for Gaussian phase-insensitive channels \cite{bib:Nat.Photonics.8.796}. On the other hand, the robustness of quantum entanglement under a noisy channel was also investigated under the same energy condition and it turns out that there exist some non-Gaussian states significantly outliving Gaussian states \cite{bib:Phys.Rev.Lett.105.100503, bib:Phys.Rev.Lett.107.130501, NJP, bib:Phys.Rev.A.90.010301}.

% summary of result
% - in CV teleportation, non-Gaussian beats Gaussian
% - similar result under same entanglement
In this paper we investigate the ultimate limit of CV quantum teleportation under energy constraint identifying an optimal quantum state that achieves the maximum fidelity. Quantum teleportation is a quantum communication protocol that not only represents a distinguishing quantum feature of entanglement but also provides a basis for many practical applications such as entanglement swapping \cite{bib:PhysRevLett.71.4287} and quantum repeaters \cite{bib:PhysRevLett.81.5932, bib:PhysRevA.59.169}. Although the original proposal of CV quantum teleportation \cite{bib:Phys.Rev.Lett.80.869} was made  using a Gaussian two-mode squeezed vacuum (TMSV) state, non-Gaussian states can also be useful under certain circumstances, with the enhanced fidelity identified in \cite{bib:Phys.Rev.A.61.032302, bib:Phys.Rev.A.65.062306, bib:Phys.Rev.A.67.032314, bib:Phys.Rev.A.73.042310}. For non-Gaussian states, the Einstein-Podolsky-Rosen (EPR) correlation is not a necessary ingredient to accomplish quantum teleportation unlike Gaussian states \cite{bib:Phys.Rev.Lett.108.030503}. 
In teleportation with multiple receivers, i.e. telecloning, the optimal fidelity is attained by non-Gaussian states \cite{bib:Phys.Rev.Lett.95.070501,bib:Phys.Rev.A.94.062318}. We investigate here the teleportation of coherent states with completely unknown displacement based on a formalism expressing the output fidelity as an expectation value of an operator $ \hat{\mathcal{F}} $. Using this operator formalism, we show that photon-number entangled states (PNESs) \cite{Lee} are optimal resources for CV teleportation and that non-Gaussian states, rigorously speaking, outperform Gaussian states under energy constraint. On the other hand, we also demonstrate that this advantage of non-Gaussian states over Gaussian states is appreciable only in the low-energy regime and it becomes negligible in the high-energy limit.

\section{CV Teleporatation fidelity in matrix representation}

% our goal
% - energy constraint
% - teleportation fidelity operator
Our goal is to find an optimal two-mode state $ \rho_{AB} $ that achieves the maximum teleportation fidelity under a given energy, or mean photon number $ \bar{n} = \tr \left[ \hat{n}_\textrm{av} \rho_{AB} \right] = \tr \left[ \frac{1}{2} \left( \hat{n}_A + \hat{n}_B \right) \rho_{AB} \right] $, where $ \hat{n}_{A(B)} $ is a photon number operator on mode $ A(B) $. We consider Braunstein-Kimble (BK) protocol \cite{bib:Phys.Rev.Lett.80.869} to teleport a coherent state with a completely unknown displacement. Let $ \hat{a}_A $, $ \hat{a}_B $, and $ \hat{a}_\textrm{in} $ be the annihilation operators corresponding to modes of Alice, Bob, and input state, respectively. The bosonic modes can also be described by their quadrature operators $ \hat{x}_j $ and $ \hat{p}_j ( j = A, B, \textrm{in} ) $, which are related to annihilation operators as $ \hat{a}_j = \frac{1}{\sqrt{2}} \left( \hat{x}_j + i\hat{p}_j \right) $. In the BK protocol, Alice superimposes the input coherent state $ \ket{\alpha_\textrm{in}} $ with her part of resource state and measures two quadratures $ \hat{x}_- \equiv \frac{1}{\sqrt{2}} \left( \hat{x}_\textrm{in} - \hat{x}_A \right) $ and $ \hat{p}_+ \equiv \frac{1}{\sqrt{2}} \left( \hat{p}_\textrm{in} + \hat{p}_A \right) $. Alice then sends her measurement outcomes $ \{ x_- , p_+ \} $ to Bob, who obtains an output state $ \rho_\textrm{out} $ by displacing his mode with the amount $ x_- + i p_+ $. 

Here we consider only pure resource states, but we show that they suffice to find the optimal states among all quantum states including mixed states. We also assume the first moments of quadratures (average amplitudes) of $\rho_{AB}$ to be zero. If nonzero, Bob can adjust his displacement to obtain the desired output state. The same level of fidelity can be achieved with zero displacement, while nonzero displacement merely increases the energy of the state without any effect on the teleportation fidelity.
As identified in \cite{bib:Phys.Rev.Lett.108.030503}, the fidelity between input and output states can be expressed by 
	\begin{equation}
F = \bra{\alpha_\textrm{in}} \rho_\textrm{out} \ket{\alpha_\textrm{in}} = \tr \left[ e^{- \hat{u}^2 - \hat{v}^2} \rho_{AB} \right] \equiv \tr \left[ \hat{\mathcal{F}} \rho_{AB} \right] ,
	\end{equation}
where $ \hat{u} \equiv \frac{1}{\sqrt{2}} \left( \hat{x}_A - \hat{x}_B \right) $ and $ \hat{v} \equiv \frac{1}{\sqrt{2}} \left( \hat{p}_A + \hat{p}_B \right) $ are the so-called EPR operators. Note that the above equation attributes the output fidelity to the property of the resource state $\rho_{AB}$.

% matrix representation
% - Block-diagonal form
Now we analyze the fidelity operator $ \hat{\mathcal{F}} $ in the Fock state basis, the eigenbasis of photon number operators. After an algebraic calculation with details in Appendix, we obtain
	\begin{align} \label{eq:Felement}
\bra{j,k} \hat{\mathcal{F}} \ket{l,m} & = f^{(d)}_{j,l} \delta_{d, k-j} \delta_{d, m-l} , \nonumber \\
\textrm{with } f^{(d)}_{j,l} & \equiv \frac{(j+l+d)!}{2^{j+l+d+1} \sqrt{j! (j+d)! l! (l+d)!}} .
	\end{align}
The Kronecker delta $ \delta_{d, k-j} \delta_{d, m-l} $ above indicates that interaction terms are nonzero only if the photon-number difference $ d $ between modes $ A $ and $ B $ is the same for the bra and ket. In other words, $ \hat{\mathcal{F}} $ can be represented in a block-diagonal form where each block corresponds to the set of states with the same photon-number difference $ d $, written as
	\begin{align} \label{eq:block}
\hat{\mathcal{F}} & = \bigoplus_{d=-\infty}^{\infty} \hat{\mathcal{F}}^{(d)} , \nonumber \\
\hat{\mathcal{F}}^{(d)} & = \sum_{j,l=0}^{\infty} f^{(d)}_{j,l} \ket{l,l+d} \bra{j,j+d} .
	\end{align}
Now let us introduce a projection operator $ \Pi^{(d)} $ that projects a state onto the subspace spanned by states with the same photon number difference $d$. Using the block-diagonal property, the fidelity can be represented as
	\begin{equation}
F = \sum_{d=-\infty}^{\infty} \tr \left[ \hat{\mathcal{F}}^{(d)} \Pi^{(d)} \rho_{AB} \Pi^{(d)} \right] .
	\end{equation}
The maximum fidelity must occur for one of the eigenstates of the operator $\hat{\mathcal{F}}$. 
Furthermore, due to the block-diagonal structure of $\hat{\mathcal{F}}$, the eigenstates of $\hat{\mathcal{F}}$ all reside in a different subspace given by $\Pi^{(d)}$, which are expressed as
	\begin{equation}
\ket{\Psi^{(d)}} = \sum_{j=0}^{\infty} c_j \ket{j,j+d} .
	\end{equation}
Our task now is to compare different $d$'s to find an optimal state and we only consider those states for $ d \ge 0 $. For $ d < 0 $, the state can be written as $ \sum_{j=0}^\infty c_j \ket{j-d,j} $, which is equivalent to the state $ \ket{\Psi^{(|d|)}} $ under permutation. Since $ f_{j,l}^{(d)} $ is symmetric under permutation between $ j $ and $ l $, these two states lead to the same fidelity as well as the same energy.

% PNES are optimal
We first find that the PNES, that is, $ \ket{\Psi^{(d)}} $ with $ d = 0 $ is optimal. Let us consider two states $ \ket{\Psi^{(0)}} $ and $ \ket{\Psi^{(d)}} $ with the same coefficients $ \{ c_j \} $. The state $ \ket{\Psi^{(0)}} $ has $ d $ fewer photons than $ \ket{\Psi^{(d)}} $ on Bob's side and the difference in fidelity is written as
	\begin{equation}
F^{(0)} - F^{(d)} = \sum_{j,l=0}^{\infty} \left( f^{(0)}_{j,l} - f^{(d)}_{j,l} \right) c_j^* c_l .
	\end{equation}
If the matrix with its element given by $ f^{(0)}_{j,l} - f^{(d)}_{j,l} $ is positive semidefinite, it implies that $ \ket{\Psi^{(0)}} $ always yields a higher fidelity with fewer resources, for any set of coefficients $ \{ c_j \} $.
	\begin{figure}[!t]
	\centering \includegraphics[width=0.8\columnwidth]{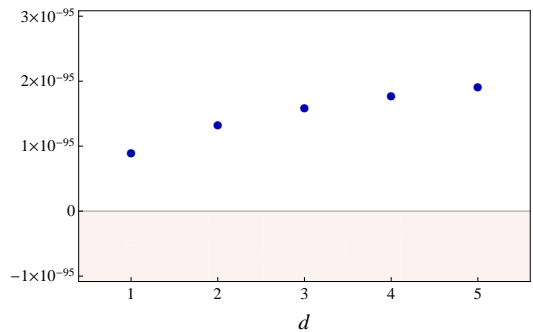}
	\caption{\label{fig:minevdf} Minimum eigenvalues of the matrices with elements given by $ f^{(0)}_{j,l} - f^{(d)}_{j,l} $. All eigenvalues lie in the positive region. }
	\end{figure}
We investigate the eigenvalues of these matrices with truncated dimension of $ j,l = 0, 1, 2, \cdots, 100 $ and their minimum eigenvalues are shown in Fig. \ref{fig:minevdf}. It can be clearly seen that for each $ d $ the minimum eigenvalue is positive, implying that the matrix is indeed positive definite. As the diagonal elements $ f^{(d)}_{j,j} $ decrease with $ j $ and $ d $ while off-diagonal elements also become negligible, increasing the truncation number does not change the positiveness of the matrices, which can also be numerically confirmed. In the next section we also present the optimal fidelity for each $ d $ explicitly, demonstrating the optimality of $ d = 0 $ again.

%Although this is not sufficient to show positive definiteness in infinite dimension, it is almost enough to show the optimality of PNES. That is because diagonal element $ f^{(d)}_{j,j} $ decreases as $ j $ or $ d $ increases and off-diagonal elements becomes smaller as they go away from diagonal terms. Also, more importantly, two-mode entangled states reside dominantly in the small photon number regime in practical situation \cite{bib:maxsqueezing}. In the next section, we also represents the optimal fidelity for each $ d $ and show that $ d = 0 $ is optimal.

\section{Optimality of non-Gaussian states}

% Lagrange multiplier method
Next we obtain the optimal fidelity by determining coefficients $ \{ c_j \} $. The optimization problem is formulated as follows:
	\begin{equation*}
\begin{aligned}
& \text{maximize} & & \bra{\Psi^{(0)}} \hat{\mathcal{F}} \ket{\Psi^{(0)}} = \sum_{j,l} f^{(0)}_{j,l} c_j^* c_l \\
& \text{subject to} & & \bra{\Psi^{(0)}} \hat{n}_\textrm{av} \ket{\Psi^{(0)}} = \sum_j j |c_j|^2 = \bar{n} .
\end{aligned}
	\end{equation*}
We may employ the Lagrange multiplier method and define a Lagrangian
	\begin{align}
\mathcal{L} & = \sum_{j,l} c_j^* c_l f^{(0)}_{j,l} - \lambda \left( \sum_j j |c_j|^2 - \bar{n} \right) \nonumber \\
& = \sum_{j,l} \left( f^{(0)}_{j,l} - \lambda j \delta_{j,l} \right) c_j^* c_l + \lambda \bar{n} \nonumber \\
& \equiv \sum_{j,l} \left( G_\lambda \right)_{j,l} c_j^* c_l + \lambda \bar{n} .
	\end{align}
Let us denote by $ g_{\lambda,\textrm{max}} $ the maximum eigenvalue of $ G_\lambda $ and by $ \ket{\Psi_{\lambda,\textrm{max}}} $ the corresponding eigenstate. 
%At this stage, it addresses the maximum of $ \mathcal{L} $, not directly the maximum of $ \langle \hat{\mathcal{F}} \rangle $ because there might exist another eigenstate with higher fidelity but larger mean photon number. However, 
For a fixed $ \lambda $, one can see that $ g_{\lambda,\textrm{max}} $ leads to the maximum fidelity $ F_\textrm{max} (\bar{n}) = g_{\lambda,\textrm{max}} + \lambda \bar{n} $ under the energy constraint $ \bar{n} = \bra{\Psi_{\lambda,\textrm{max}}} \hat{n}_\textrm{av} \ket{\Psi_{\lambda,\textrm{max}}} $ as follows. Let us consider the states other than $ \ket{\Psi_{\lambda,\textrm{max}}} $ but satisfying the same energy constraint. Those states can generally be written as a superposition of different eigenstates of $ G_\lambda $. Even though some components of the superposition might yield higher fidelity, the net fidelity written as $ \langle G_\lambda \rangle + \lambda \bar{n} $ cannot be larger than the optimal fidelity since the first term $ \langle G_\lambda \rangle $ must be smaller than $ g_{\lambda,\textrm{max}} $ while the second term $ \lambda \bar{n} $ remains constant. This argument is also valid for mixed states, which implies that a pure state $ \ket{\Psi_{\lambda,\textrm{max}}} $ becomes optimal among all quantum states.

Notably, $ \lambda $ represents the rate at which $ F_\textrm{max} (\bar{n})$ changes with respect to $ \bar{n} $, that is, $ \lambda = dF_\textrm{max}/d\bar{n} $ because $ F_\textrm{max} (x) - \lambda x $ has an extremal point at $ x = \bar{n} $. We may estimate the range of $ \lambda $ to consider for a given $ \bar{n} $. For example, for large $ \bar{n} $, we expect $ \lambda $ to be close to zero because the fidelity asymptotically converges to 1. With $ \lambda $ increasing, $ \bar{n} $ of the optimal state becomes smaller. This way, the optimal eigenstate of $ G_\lambda $ by varying $\lambda$ can give an optimal fidelity for a different energy $ \bar{n} $. 

% Optimality of TMSV in the Gaussian regime
Actually $ G_\lambda $ is a matrix representation of the operator $ \hat{\mathcal{F}}^{(0)} - \lambda \hat{n}_\textrm{av} $ in the basis $ \ket{j,j} $. This operator is non-Gaussian, thus proving that Gaussian photon-number entangled states, TMSVs, are not optimal.
Before moving on, we briefly discuss the optimality of TMSVs within the Gaussian regime. Because we assume zero displacement, Gaussian states are fully described by their second moments, represented as a covariance matrix. For two-mode pure states, a standard form of covariance matrix corresponds to TMSVs. On the other hand, the upper bound of teleportation fidelity for Gaussian states was derived in terms of the lowest symplectic eigenvalue $ \nu $ of partial transposed state \cite{bib:Phys.Rev.A.78.062340}, which is directly related to the entanglement negativity \cite{bib:Phys.Rev.A.65.032314}. The upper bound is reached if the state is symmetric on both sides and the TMSVs thus achieve the upper bound. Any other states that are not in the standard form can be generated by applying local symplectic operation, phase rotation, and local squeezing, on the TMSV. However these operation do not change $ \nu $ and never decrease the mean photon number. Therefore TMSV has the minimum mean photon number among states with the same $ \nu $ and thus achieves the maximum fidelity under the energy constraint. The maximum fidelity achieved by the TMSV is given by $ F^{(G)} = 1/(1+e^{-2r}), $ with $ r $ a squeezing parameter and the energy constraint $ \bar{n} = \sinh^2 r $.

% result
While it is intractable to analytically obtain the eigenvalues of $ G_\lambda $ in infinite dimension, it is sufficient to investigate a truncated matrix with $ j,l = 0, 1, 2, \cdots, n_\textrm{trunc} $ with $ n_\textrm{trunc} \gg \bar{n} $. We numerically find the eigenvalues and the eigenvectors by varying $ \lambda $ using a sufficiently large truncation number $ n_\textrm{trunc} = 100 $. For each $ \lambda $, we obtain its maximum eigenvalue and the corresponding eigenvector with mean photon number $ \bar{n} $.
	\begin{figure}[!t]
	\centering \includegraphics[width=0.9\columnwidth]{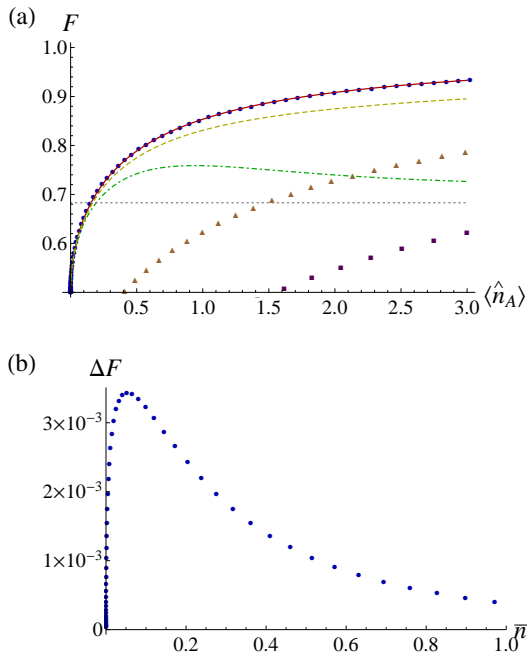}
	\caption{\label{fig:fidelity} (a) Optimal fidelity achieved by PNESs (blue circles) compared to the fidelity by TMSV (red curve) and by $ \ket{\Psi^{(d)}} $ with $ d = 1 \textrm{ (brown triangles)}$ and 2 (purple squares) against energy $\langle \hat{n}_A \rangle $. We also plot the fidelity attained by other non-Gaussian states: TMCs (green dot-dashed curve) and PSSVs (yellow dashed curve). The gray dotted line represents the no-cloning bound $ F_\textrm{nc} \approx 0.6826 $. (b) Difference in the optimal fidelity between non-Gaussian states and Gaussian states against energy $\bar n$. }
	\end{figure}
In Fig. \ref{fig:fidelity}(a) we show the optimal fidelity achieved by PNESs compared to the fidelity of the TMSV and of $ \ket{\Psi^{(d)}} $ with $ d = 1,2 $. First, the optimal fidelity attained with $ d = 0 $ clearly beats the fidelity with a nonzero $ d $. Note that we represent the fidelity against $ \langle \hat{n}_A \rangle $, not $ \langle \hat{n}_\textrm{av} \rangle = \bar{n} $. The states $ \ket{\Psi^{(d)}} $ have $ d $ more photons on Bob's side, but they attain a smaller fidelity. When $ d $ is nonzero, the fidelity can not even reach the classical bound $ F_\textrm{cl} = 1/2 $, above which the quantum nature of teleportation is demonstrated, in the small photon regime.

% - slight improvement by non-Gaussian states
On the other hand, Gaussian states, TMSVs, show near-optimal fidelity. The difference in fidelity between TMSVs and the optimal PNES is shown in Fig. \ref{fig:fidelity}(b). Employing non-Gaussian states takes a slight improvement, especially in the small photon regime, particularly $ \Delta F \simeq 0.0035 $ at $ \bar{n} \simeq 0.052 $. To obtain the same fidelity using TMSVs, $ \bar{n} \simeq 0.055 $ is required, about $ 6 \% $ more resources.

% photon number distribution
	\begin{figure}[!t]
	\centering \includegraphics[width=0.8\columnwidth]{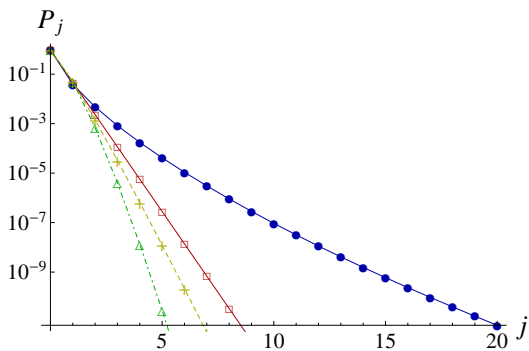}
	\caption{\label{fig:distribution} Photon-number distribution $ P_j = | c_j |^2 $ of the optimal PNESs (blue circles) and of TMSVs (red squares), with the same mean photon number $ \bar{n} \simeq 0.052 $. Also the photon-number distribution of TMCs (green triangles) and of PSSVs (yellow pluses) are shown.}
	\end{figure}
To see the exact form of the optimal state, we display its photon-number distribution in Fig. \ref{fig:distribution}, together with the distribution of TMSVs. %The distribution is almost the same at $ j = 0, 1 $, but 
The distribution of TMSVs decays much faster than that of the optimal PNES, which has a rather long-tail distribution slightly affecting the teleportation fidelity despite low probabilities. Photon-number statistics are usually characterized by Mandel's $ Q $ factor, defined as $ Q = \langle (\Delta n)^2 \rangle / \langle n \rangle - 1 $. To exhibit a long-tail distribution, a larger $ Q $ factor is required.

We compare the optimal PNES with other types of non-Gaussian PNESs that have been intensively studied: (i) two-mode coherently correlated states (TMCs) with $ c_j \propto x^j / j! $ \cite{bib:Phys.Rev.Lett.57.827} and (ii) photon-subtracted squeezed states (PSSVs) with $ c_j \propto (j+1) x^{j+1} $ \cite{bib:Phys.Rev.A.61.032302}. Similar to PSSV, photon-added squeezed states (PASVs) have also been investigated but they are not of interest here because the mean photon number of PASVs is always greater than 1 while the advantage of non-Gaussian states in CV teleportation appears in the small photon-number regime. It was found that PSSVs make an improvement in teleportation fidelity compared to TMSVs with the same initial squeezing \cite{bib:Phys.Rev.A.61.032302}. However, both PSSVs and TMCs fail to beat TMSVs under the same energy as shown in Fig. \ref{fig:fidelity}(a). In Fig. \ref{fig:distribution}, we show the photon number distribution of TMCs and of PSSVs, which are fast decaying in contrast to the long-tail distribution of the optimal PNES. The $ Q $ factor becomes smaller as the distribution decays faster and such a small $ Q $ factor may be used in other applications because it reveals a nonclassical feature. %However, for CV quantum teleportation, a high Q factor is desired.

\section{Channel loss analysis}

In practical situations, entangled states are distributed over two distant parties hardly without any disturbance. Practically important to consider is a loss channel with transmittance $ \eta $. We examine if the optimal PNESs beat TMSVs after dissipated by the same channel loss. We assume that each side of two-mode states undergoes the identical channel with the same $ \eta $. Note that if two different input states have the same mean photon number $ \bar{n} $, the output states also have the same mean photon number $ \eta \bar{n} $.

	\begin{figure}[!t]
	\centering \includegraphics[width=0.8\columnwidth]{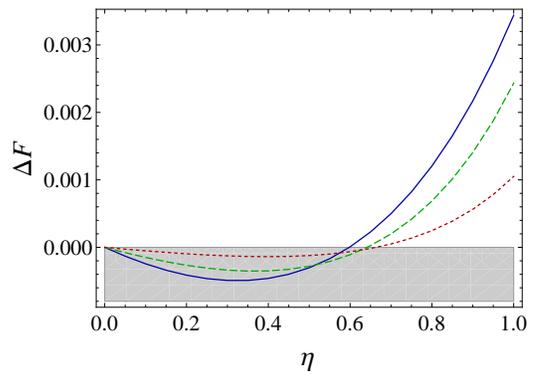}
	\caption{\label{fig:loss} Fidelity difference between Gaussian states and non-Gaussian states under the channel loss of transmittance $ \eta $. The initial mean photon number $ \bar{n} $ is chosen as 0.052 (blue solid curve), 0.204 (green dashed curve), and 0.515 (red dotted curve). The gray region represents that Gaussian states beat non-Gaussian states. }
	\end{figure}
In Fig. \ref{fig:loss} we plot the fidelity difference between Gaussian states and non-Gaussian states after dissipated by the same channel loss. The non-Gaussian states maintain their advantage over Gaussian states for a quite moderate value of transmittance $ \eta \gtrsim 0.7 $. If the initial difference at $ \eta = 1 $ is larger, the state endures more loss. In particular, the state that shows the maximum $ \Delta F $ for $ \bar{n} \simeq 0.052 $ keeps its advantage over the Gaussian state to $ \eta \gtrsim 0.6 $.

\section{Conclusion}

% summary
% entanglement constraint
We have investigated CV quantum teleportation to find an ultimate limit of fidelity under energy constraint and identified non-Gaussian photon-number entangled states as rigorously optimal. The optimal states have a long-tail distribution in contrast to so-far widely considered non-Gaussian states such as TMCs or PSSVs. While those non-Gaussian states make a slight improvement over Gaussian states, the enhancement becomes negligible in the large-energy limit. In other words, Gaussian states, particularly TMSVs, are near optimal in a wide range. This can be contrasted with the robustness of entanglement against a Gaussian noisy channel for which some non-Gaussian states manifest a significant advantage over Gaussian counterparts \cite{bib:Phys.Rev.Lett.105.100503, bib:Phys.Rev.Lett.107.130501, NJP, bib:Phys.Rev.A.90.010301}. In a sense, however, the teleportation fidelity studied here can be regarded as a single specific entanglement criterion, whereas the correlation studied in \cite{bib:Phys.Rev.Lett.105.100503, bib:Phys.Rev.Lett.107.130501, NJP, bib:Phys.Rev.A.90.010301} addresses the whole of entanglement.

Although we have discussed our problem with energy constraint, we can make the same argument by considering entanglement as a resource. For pure states, entanglement is generally measured by von Neumann entropy and the TMSV achieves the maximum under the same energy constraint. The non-Gaussian PNESs we find optimal achieve higher fidelity with the same energy and thus with smaller entanglement. This is clear evidence that non-Gaussian states beat Gaussian states with the same entanglement. % although they may not be optimal states.

The present work provides the fundamental upper bound of CV teleportation fidelity, that is, to what extent teleportation fidelity can be achieved within a given energy. This becomes a practically relevant issue when one generates a two-mode entangled state by supplying energy starting from a vacuum state. On the other hand, there are alternative routes to generate entangled states. For example, PSSVs can be simply generated by single photon counting after a beamsplitter interaction, although nondeterministic. In this case, one may consider other constraint of resource rather than the energy constraint, e.g., technical demands of quantum operations, which can be another interesting line of study in the future.  %Other than teleportation fidelity, there are different quantities associated with CV entanglement, which were investigated with the same covariance condition sometimes exhibiting extremality properties of Gaussian states \cite{bib:Phys.Rev.Lett.96.080502}.Under the eneregy constraint, we investigate various quantities such as entanglement negativity and Bell inequality in our future work \cite{bib:FutureWork}.

\section*{acknowledgement}
We acknolwedge the support by an NPRP grant 7-210-1-032 from Qatar National Research Fund.

\onecolumngrid
\appendix

\section{\label{sec:matrix}Fidelity operator $ \hat{\mathcal{F}} $ in Fock state basis}

We present the details of the matrix element of the fidelity operator $ \hat{\mathcal{F}} = e^{- \hat{u}^2 - \hat{v}^2} = e^{- (\hat{a}_B^\dagger - \hat{a}_A) (\hat{a}_B - \hat{a}_A^\dagger)} $ in the Fock state basis. Using the Baker-Campbell-Hausdorff relation $ e^{-\hat{A}} \hat{B} e^{\hat{A}} = \hat{B} - \left[ \hat{A}, \hat{B} \right] + \frac{1}{2!} \left[ \hat{A}, \left[ \hat{A}, \hat{B} \right] \right] - \frac{1}{3!} \left[ \hat{A}, \left[ \hat{A}, \left[ \hat{A}, \hat{B} \right] \right] \right] + \cdots $ with the commutation relations $ \left[ (\hat{a}_B^\dagger - \hat{a}_A) (\hat{a}_B - \hat{a}_A^\dagger) , \hat{a}_A^\dagger \right] = \hat{a}_A^\dagger - \hat{a}_B $ and  $ \left[ (\hat{a}_B^\dagger - \hat{a}_A) (\hat{a}_B - \hat{a}_A^\dagger) , \hat{a}_B^\dagger \right] = \hat{a}_B^\dagger - \hat{a}_A $, we have
	\begin{align}
\bra{j,k} \hat{\mathcal{F}} \ket{l,m} & = \frac{1}{\sqrt{j!k!l!m!}} \bra{0,0} \hat{a}_A^j \hat{a}_B^k e^{- (\hat{a}_B^\dagger - \hat{a}_A) (\hat{a}_B - \hat{a}_A^\dagger)} ( \hat{a}_A^\dagger )^l ( \hat{a}_B^\dagger )^m \ket{0,0} \nonumber \\
& = \frac{1}{\sqrt{j!k!l!m!}} \bra{0,0} \hat{a}_A^j \hat{a}_B^k \hat{a}_B^l \hat{a}_A^m e^{- (\hat{a}_B^\dagger - \hat{a}_A) (\hat{a}_B - \hat{a}_A^\dagger)} \ket{0,0} \nonumber \\
& = \frac{1}{\sqrt{j!k!l!m!}} \bra{0,0} \hat{a}_A^{j+m} \hat{a}_B^{k+l} \hat{\mathcal{F}} \ket{0,0} .
	\end{align}
The operator $ \hat{\mathcal{F}} $ can be expanded in the terms of operators $ \hat{a}_B^\dagger \hat{a}_B $, $ \hat{a}_B^\dagger \hat{a}_A^\dagger $, $ \hat{a}_A \hat{a}_B $, and $ \hat{a}_A \hat{a}_A^\dagger $, for which the photon numbers on both modes increase and decrease simultaneously, or remain unchanged. Therefore, the last line of the above equation becomes nonzero only if $ j+m = k+l $. This is why $ \hat{\mathcal{F}} $ can be written in a block-diagonal form as Eq. (\ref{eq:block}). From now on let us define $ k-j = m-l \equiv d $ and $ j+m = k+l \equiv s $ for simplicity.

Returning to $ \hat{\mathcal{F}} = e^{- \hat{u}^2 - \hat{v}^2} $, $ \hat{u} $ and $ \hat{v} $ can be generated by mixing quadratures of mode $ A $ and $ B $ at a 50/50 beamsplitter, that is, $ \hat{U}_\textrm{BS} \hat{x}_A \hat{U}_\textrm{BS}^\dagger = \hat{u} $ and $ \hat{U}_\textrm{BS} \hat{p}_B \hat{U}_\textrm{BS}^\dagger = \hat{v} $. Then nonzero terms of $ \bra{j,k} \hat{\mathcal{F}} \ket{l,m} $ can be written as
	\begin{align} \label{eq:Felement2}
\bra{j,j+d} \hat{\mathcal{F}} \ket{l,l+d} & = \frac{1}{\sqrt{j!(j+d)!l!(l+d)!}} \bra{0,0} \hat{a}_A^s \hat{a}_B^s \hat{U}_\textrm{BS} e^{-\hat{x}_A^2 - \hat{p}_B^2} \hat{U}_\textrm{BS}^\dagger \ket{0,0} \nonumber \\
& = \frac{1}{\sqrt{j!(j+d)!l!(l+d)!}} \bra{0,0} \left( \frac{\hat{a}_A + \hat{a}_B}{\sqrt{2}} \right)^s \left( \frac{-\hat{a}_A + \hat{a}_B}{\sqrt{2}} \right)^s e^{-\hat{x}_A^2 - \hat{p}_B^2} \ket{0,0} \nonumber \\
& = \frac{1}{2^s\sqrt{j!(j+d)!l!(l+d)!}} \sum_{t_1=0}^{s} \sum_{t_2=0}^{s} (-1)^{t_2} \binom{s}{t_1} \binom{s}{t_2} \bra{0,0} \hat{a}_A^{t_1+t_2} \hat{a}_B^{2s-t_1-t_2} e^{-\hat{x}_A^2 - \hat{p}_B^2} \ket{0,0} \nonumber \\
& = \sum_{t_1=0}^{s} \sum_{t_2=0}^{s} \frac{(-1)^{t_2}}{2^s} \sqrt{\frac{(t_1+t_2)! (2s-t_1-t_2)!}{j!(j+d)!l!(l+d)!}} \binom{s}{t_1} \binom{s}{t_2} \bra{t_1+t_2} e^{-\hat{x}_A^2} \ket{0}_A \bra{2s-t_1-t_2} e^{-\hat{p}_B^2} \ket{0}_B .
	\end{align}
The brakets in the last line can be evaluated using the quadrature representation of the Fock state $ \braket{x_\theta}{n} = \frac{1}{\sqrt{\pi^{1/2} 2^n n!}} e^{-x_\theta^2/2} H_n(x_\theta) e^{-in\theta} $, with $ \hat{x}_\theta = \hat{x}\cos\theta + \hat{p}\sin\theta $ and $ H_n(x_\theta) $ the Hermite polynomial of order $ n $. It yields
	\begin{align} \label{eq:Hermiteint}
\bra{n} e^{-x_\theta^2} \ket{0} & = \frac{1}{\sqrt{\pi 2^n n!}} \int_{-\infty}^\infty dx e^{-2x^2} H_n(x) H_0(x) e^{in\theta} \nonumber \\
& = \begin{cases}
\frac{(-1)^{n/2}\sqrt{n!}}{2^{n+1/2}\left(\frac{n}{2}\right)!} e^{in\theta} & \textrm{if $ n $ is even} \\
0 & \textrm{if $ n $ is odd .}
\end{cases}
	\end{align}
Substituting Eq. (\ref{eq:Hermiteint}) into Eq. (\ref{eq:Felement2}) leads to
	\begin{equation} \label{eq:Felement4}
(\ref{eq:Felement2}) = \sum_{t_1=0}^{s} \sum_{t_2=0}^{s} \frac{(-1)^{\frac{t_1-t_2}{2}}}{2^{3s+1}\sqrt{j!(j+d)!l!(l+d)!}} \frac{(t_1+t_2)!(2s-t_1-t_2)!}{(\frac{t_1+t_2}{2})!(s-\frac{t_1+t_2}{2})!} \binom{s}{t_1} \binom{s}{t_2} ,
	\end{equation}
where the summation runs over all pairs of $ \{ t_1 , t_2 \} $ with $ t_1+t_2 $ even, or equivalently $ t_1-t_2 $ even. The summation over odd $ t_1-t_2 $ is eliminated due to the odd symmetry under permutation between $ t_1 $ and $ t_2 $. By introducing variables $ t_+ = \frac{t_1+t_2}{2} $ and $ t_- = \frac{t_1-t_2}{2} $, we rewrite the Eq. (\ref{eq:Felement4}) as
	\begin{equation} \label{eq:Felement5}
\sum_{t_+=0}^{s} \sum_{t_-=-t_+}^{t_+} \frac{(-1)^{t_-}}{2^{3s+1}\sqrt{j!(j+d)!l!(l+d)!}} \frac{(2t_+)!(2s-2t_+)!}{t_+!(s-t_+)!} \binom{s}{t_++t_-} \binom{s}{t_+-t_-} .
	\end{equation}
To evaluate the summation over $ t_- $, we use the binomial expansion of $ (1+y^2)^s $ as
	\begin{align}
\sum_{t=0}^s \binom{s}{t} y^{2t} & = (1+y^2)^s = (1+iy)^s (1-iy)^s \nonumber \\
& = \sum_{t_1=0}^{s} \sum_{t_2=0}^{s} i^{t_1-t_2} \binom{s}{t_1} \binom{s}{t_2} y^{t_1+t_2} \nonumber \\
& = \sum_{t_+=0}^s \sum_{t_-=-t_+}^{t_+} (-1)^{t_-} \binom{s}{t_++t_-} \binom{s}{t_+-t_-} y^{2t_+} , \nonumber
	\end{align}
therefore
	\begin{equation} \label{eq:binomsum}
\binom{s}{t_+} = \sum_{t_-=-t_+}^{t_+} (-1)^{t_-} \binom{s}{t_++t_-} \binom{s}{t_+-t_-} .
	\end{equation}
Now, with the relation (\ref{eq:binomsum}), Eq. (\ref{eq:Felement5}) becomes
	\begin{equation}
(\ref{eq:Felement5}) = \frac{s!}{2^{3s+1}\sqrt{j!(j+d)!l!(l+d)!}} \sum_{t_+=0}^{s} \binom{2t_+}{t_+} \binom{2s-2t_+}{s-t_+} .
	\end{equation}
The summation over $ t_+ $ can be evaluated as \cite{bib:combinatorial}
	\begin{equation}
\sum_{t_+=0}^{s} \binom{2t_+}{t_+} \binom{2s-2t_+}{s-t_+} = 2^{2s} .
	\end{equation}
Finally, we have
	\begin{equation}
\bra{j,j+d} \hat{\mathcal{F}} \ket{l,l+d} = \frac{(j+l+d)!}{2^{j+l+d+1}\sqrt{j!(j+d)!l!(l+d)!}} ,
	\end{equation}
which proves Eq. (\ref{eq:Felement}).

\end{document}